\documentclass[submission,copyright,sharealike]{eptcs}

\usepackage[]{sosy-paper}

\usepackage[strings]{underscore} 
\usepackage{breakurl}

\newcommand{\locationcpa}{{LocationCPA}\xspace}
\newcommand{\callstackcpa}{{CallstackCPA}\xspace}
\newcommand{\threadingcpa}{{ThreadingCPA}\xspace}
\newcommand{\pthreadcreate}{{$pthread\_create$}\xspace}
\newcommand{\pthreadjoin}{{$pthread\_join$}\xspace}

\title{A Light-Weight Approach for Verifying Multi-Threaded Programs with CPAchecker}

\author{Dirk Beyer \institute{LMU Munich, Germany}
    \and Karlheinz Friedberger \institute{University of Passau, Germany}}

\def\titlerunning{A Light-Weight Approach for Verifying Multi-Threaded Programs with CPAchecker}
\def\authorrunning{D.\ Beyer \& K.\ Friedberger}
\providecommand{\event}{MEMICS 2016}

\begin{document}
\maketitle
\begin{abstract}
	Verifying multi-threaded programs is becoming more and more important,
because of the strong trend to increase the number of processing units per CPU socket.
We introduce a new configurable program analysis
for verifying multi-threaded programs with a bounded number of threads.
We present a simple and yet efficient implementation as component of the existing 
program-verification framework \cpachecker.
While \cpachecker is already competitive on a large benchmark set of sequential verification tasks,
our extension enhances the overall applicability of the framework.
Our implementation of handling multiple threads is orthogonal to the abstract domain of the data-flow analysis,
and thus, can be combined with several existing analyses in \cpachecker,
like value analysis, interval analysis, and BDD analysis.
The new analysis is modular and can be used, for example, 
to verify reachability properties as well as to detect deadlocks in the program.
This paper includes an evaluation of the benefit of some optimization steps
(e.g., changing the iteration order of the reachability algorithm or applying partial-order reduction)
as well as the comparison with other state-of-the-art tools for verifying multi-threaded programs.

\end{abstract}

\section{Introduction}

Program verification has successfully been applied to programs to find errors in applications.
There exist many approaches to verify single-threaded programs (cf.~SV-COMP for an overview~\cite{SVCOMP16}),
and several of them are already implemented in the
open-source program-verification framework \cpachecker~\cite{CPACHECKER, CPACHECKER-COMP15}.
For multi-threaded programs a new dimension of complexity has to be taken into account:
the verification tool has to efficiently analyze all possible thread interleavings.
%
\cpachecker did not support the analysis of multi-threaded programs for a long time.
Our work focuses on a new, simple configurable program analysis
that reuses several existing components of the framework.
The approach is sound and can be combined with several steps of optimization
to achieve an efficient analysis for multi-threaded programs.

Our analysis is based on a standard state-space exploration
using a given control-flow automaton that represents the program.
For a program state with several active threads,
we compute the succeeding program state for each of those threads,
i.e. basically we compute every possible interleaving of the threads.
%
The approach is orthogonal to other data-flow based analyses in \cpachecker,
thus it can be combined with algorithms like CEGAR~\cite{ClarkeCEGAR}
and analyze an potentially infinite state space.

\subsubsection{Related Work}
%
A prototypical version of our analysis was already applied
for the category of concurrent programs during the SV-COMP'16~\cite{SVCOMP16}.
Due to some unsupported features and
missing parts of the optimization that where implemented later,
the score in this category was low at that time.
The experimental results that we report show that the
current version of the implementation performs much better.


Just like several other tools~\cite{SPIN, SATABS, CordeiroF11ESBMC},
we explore possible interleavings of different thread executions
and our optimization methods include partial order reduction~\cite{Godefroid96POR}.
%
In contrast to verification techniques for multi-threaded programs like
constraint-based representation~\cite{THREADER}
that limits the domain to Horn clauses and predicate abstraction
or sequentialization~\cite{CSEQ-COMP13, Inverso15LazyCSeq}
that transforms the program on source-code level before starting the analysis,
our approach computes the interleaving of threads on-the-fly
and is independent from the applied analysis.
This makes it possible
to integrate our approach easily with data-flow analyses of different abstract domains,
such as value analysis~\cite{CPAexplicit} and BDD analysis~\cite{CPABDD}.

\section{Analysis of Multi-Threaded Programs in \cpachecker}

The following section provides an overview of some basic concepts and definitions used for our approach.
We describe the program representation
and the details of our configurable program analysis.

\subsection{Program Representation}

A program is represented by a \emph{control-flow automaton} (CFA) $A = (L,l_{0},G)$, which consists of
a set $L$ of program locations (modeling the program counter),
a set $G \subseteq L \times {Ops} \times L$
(modeling the control flow with assignment and assumption operations from $Ops$), and
an initial program location $l_{0}$ (entry point of the main function).



Let $V$ be the set of variables in the program.
The \emph{concrete data state for a program location} assigns a value to each variable from the set $V$;
the set~$C$ contains all concrete data states.
For every edge $g \in G$, the transition relation is defined by 
$\stackrel{g}{\rightarrow} \subseteq C \times \lbrace g \rbrace \times C$.
The union of all edges defines the complete transfer relation
$\rightarrow = \bigcup_{g \in G} \stackrel{g}{\rightarrow}$.
%
If there exists a chain of concrete data states $\langle c_0, c_1, ... , c_n \rangle$ 
with $\forall c_i:$ there exists a program location~$l_i$ for which $c_i$ is a concrete data state
and $\forall i : 1 \leq i \leq n \Rightarrow \forall i : 1 \leq i \leq n \Rightarrow \exists g: c_{i-1} \stackrel{g}{\rightarrow} c_i \land (l_{i-1}, g, l_i) \in G$,
then the state $c_n$ is \emph{reachable} from $c_0$ for~$l_0$.


Our analysis is a reachability analysis
and unrolls the program lazily~\cite{LazyAbstraction} into an \emph{abstract reachability graph} (ARG)~\cite{BLAST}.
The ARG is a directed acyclic graph
that consists of abstract states (representing the abstract program state, 
e.g., including program location and variable assignments)
and edges modeling the transfer relation that leads from one abstract state to the next one.
%



\subsection{\threadingcpa}

\cpachecker is based on the concept of \emph{configurable program analysis} (CPA)~\cite{CPA}.
Thus, different aspects of a program are analyzed by different components (denoted as CPAs).
A default analysis in \cpachecker~\cite{CPACHECKER} uses
the \locationcpa to track the program location (program counter)
and the \callstackcpa to track call stacks
(function calls and their corresponding return location in the CFA).
Thus, for the analysis of sequential programs, each abstract state that is reached during an analysis consists
of exactly one program location and one call stack.

For the analysis of multi-threaded programs we have developed a new \threadingcpa
that replaces both the \locationcpa and the \callstackcpa
and explores the state space of a multi-threaded program on-the-fly.
The benefit of the \threadingcpa is that it is able to track several program locations (one per thread)
together with their call stacks (also one per thread).
For simplicity of the definition we ignore the handling of call stacks in the next section.
The reader can simply assume that for each program location there is also a call stack.
The \threadingcpa has to handle multiple call stacks (one per thread),
whereas the \callstackcpa only handles a single call stack.

The definition of the \threadingcpa 
$\mathbb{T} = (D_\mathbb{T}, \rightsquigarrow_\mathbb{T}, {merge}_\mathbb{T}, {stop}_\mathbb{T})$
follows the structure of a configurable program analysis:

\newcommand{\threadDomain}{\mathcal{T}}

\noindent\textbf{Domain:}
    The abstract domain $D_\threadDomain = (C,\threadDomain,[[\cdot]])$ is a triple of
    the set $C$ of concrete states,
    the flat semi-lattice $\threadDomain = (T, \sqsubseteq, \sqcup, \top)$, and
    the concretization function $[[\cdot]]:\threadDomain \rightarrow 2^C$.
%
    Let $I$ be the set of all possible thread identifiers,
    e.g., a set of names used to identify threads in the program.
    The type of abstract states $T : I \pto \mathcal{L}$ consists of
    all assignments of thread identifiers $t \in I$
    to program locations $l \in \mathcal{L} = L \cup \{ \top_L \}$.
    The special program location $\top_L$ represents an unknown program location.
    The top element $\top \in T$, with $\top(t) = \top_L$ for all $t \in I$,
    is the abstract state
    that holds no specific program location for any thread identifier.
    Each abstract threading state $s \in T$ is represented by
    the assignments $\{ t_1 \mapsto l^{t_1}, t_2 \mapsto l^{t_2}, ...\}$
    of thread identifiers to their current program location.
    The partial order $\sqsubseteq$ induces a semi-lattice for the abstract states.
    The join operator $\sqcup$ yields the least upper bound of given abstract states.
    The top element $\top$ of the semi-lattice is defined as $\top = \sqcup T$.

\noindent\textbf{Merge:}
    The \threadingcpa uses the merge operator ${merge}_{sep}$,
    which does not combine different elements.

\noindent\textbf{Stop:}
    The \threadingcpa uses the termination operator ${stop}_{sep}$,
    which defines coverage only in case of equal abstract states.

\noindent\textbf{Transfer:}
    The transfer relation $\rightsquigarrow_\mathbb{T}$ determines the syntactic successor for the current state
    and is based on the transfer relation of the \locationcpa.
    The implementation is simple:
    The transfer relation returns all possible successors for all threads that are active in an abstract state,
    i.e., it applies the transfer relation of the \locationcpa for each active thread.
    Additionally, thread-management-related operations are included,
    such that creating or joining threads
    (when calling \pthreadcreate or \pthreadjoin) is defined.
    It is in theory sufficient to only handle these two function calls,
    because other thread-related function calls do not change
    the number of threads or the progress of the state-space exploration.
    The transfer relation $\rightsquigarrow_\mathbb{T}$ has the transfer 
    $s \stackrel{g}{\rightsquigarrow} s'$ for two abstract states
    $s = \{ t_1 \mapsto l^{t_1}, t_2 \mapsto l^{t_2}, ..., t_N \mapsto l^{t_N} \}$ and
    $s' = \{ t_1 \mapsto l'^{t_1}, t_2 \mapsto l'^{t_2}, ..., t_N \mapsto l'^{t_M} \}$ and $g=(l^{t_i}, op, l'^{t_i})$ if
    \begin{enumerate}
        \item the operation $op$ matches the \pthreadcreate statement
            for $t_i$ that is in program location $l^{t_i}$
            and creates a new thread $t_{new}$ starting from a CFA node $l_0^{t_{new}} \in L$:
\vspace{-.8mm}
            $$s' = s
                    \setminus \{ t_i \mapsto l^{t_i} \}
                    \cup \{ t_{new} \mapsto l_0^{t_{new}} \}
                    \cup \{ t_i \mapsto l'^{t_i} \}$$
\vspace{-.8mm}
            i.e., an existing thread $t_i$ matches the program location $l^{t_i}$
            and moves along the edge $g$ towards program location $l'$,
            and the initial program location $l_0^{t_{new}}$ of the new thread $t_{new}$
            is added to the current abstract state.

        \item the operation $op$ matches the \pthreadjoin statement
            for $t_i$ that is in program location $l^{t_i}$
            and waits for a thread $t_{exit}$ to exit,
            $t_{exit}$ exits at program location $l_E^{t_{exit}}$,
            and $t_{exit} \mapsto l_E^{t_{exit}} \in s$:
\vspace{-.8mm}
            $$s' = s
                    \setminus \{ t_i \mapsto l^{t_i} \}
                    \setminus \{ t_{exit} \mapsto l_E^{t_{exit}}\}
                    \cup \{ t_i \mapsto l'^{t_i} \}$$
\vspace{-.8mm}
            i.e., an existing thread $t_i$ matches the program location $l^{t_i}$
            and moves along the edge $g$ towards program location $l'^{t_i}$,
            and the program location $l_E^{t_{exit}}$ of the thread $t_{exit}$
            is removed from the current abstract state,
            if the thread $t_{exit}$ has already been at this program location.

        \item otherwise, if the operation $op$ is not related to thread management:
\vspace{-.8mm}
            $$s' = s
                    \setminus \{ t_i \mapsto l^{t_i} \}
                    \cup \{ t_i \mapsto l'^{t_i} \}$$
\vspace{-.8mm}
            i.e., thread $t_i$ matches the program location $l^{t_i}$
            and moves along the edge towards $l'^{t_i}$.
            
    \end{enumerate}        


For a basic analysis for multi-threaded programs
the handling of the operations \pthreadcreate and \pthreadjoin is sufficient.
Additional thread management like mutex locks (details in Section~\ref{opt-mutex-locks})
can be applied on top of this transfer relation.
We assume C statements as atomic statements,
i.e., interleaving of threads is considered to happen on statement level
(matching the encoding of the program as CFA).
This might be insufficient for real-world programs,
but is good enough for several examples and
in theory the CFA could be inflated 
with read and write operations for memory registers.

\subsection{Example}

\begin{figure}
  \begin{minipage}[b]{.48\textwidth}
    \centering
    \scalebox{.9}{
\begin{lstlisting}[language=C, firstline=12]
pthread_t id1, id2;
int i=1, j=1;

void main() {
  pthread_create(&id1, 0, t1, 0);
  pthread_create(&id2, 0, t2, 0);

  pthread_join(id1, 0);
  pthread_join(id2, 0);

  assert(j <= 8);
}

void t1() {
  i+=j;
  i+=j;
}

void t2() {
  j+=i;
  j+=i;
}
\end{lstlisting}
    }
    \caption{Program with concurrent threads}
    \label{example-code}
  \end{minipage}
  \hfill
  \begin{minipage}[b]{.48\textwidth}
    \centering
    \scalebox{.8}{


\tikzset{
  N/.style = { shape=circle, draw=black, inner sep=0pt, minimum size=12pt }
}
\begin{tikzpicture}[x=1cm,y=.8cm, node distance=.8cm]
  \def\pthread{pthread\_}
  \foreach \x in {0,1,2,3,4,5,6,7} {
    \node[N] at (0,-\x) (\x) {\x};
  }
  \path[->, thick] (0) edge node[right] {pthread\_t id1, id2;} (1);
  \path[->, thick] (1) edge node[right] {int i=1; j=1} (2);
  \path[->, thick] (2) edge node[right] (create) {{\pthread}create(\&id1, 0, t1, 0);} (3);
  \path[->, thick] (3) edge node[right] {{\pthread}create(\&id2, 0, t2, 0);} (4);
  \path[->, thick] (4) edge node[right] {{\pthread}join(\&id1, 0);} (5);
  \path[->, thick] (5) edge node[right] {{\pthread}join(\&id2, 0);} (6);
  \path[->, thick] (6) edge node[right] {assert(j<=8);} (7);

  \node[N] (10) at (0,-8.5) {A};
  \node[N, below of=10] (11) {B};
  \node[N, below of=11] (12) {C};
  \path[->, thick, ForestGreen, text=black] (10) edge node[right] (sum1) {i+=j;} (11);
  \path[->, thick, ForestGreen, text=black] (11) edge node[right] {i+=j;} (12);

  \node[N] (20) at (3,-8.5) {X};
  \node[N, below of=20] (21) {Y};
  \node[N, below of=21] (22) {Z};
  \path[->, thick, Red, text=black] (20) edge node[right] (sum2) {j+=i;} (21);
  \path[->, thick, Red, text=black] (21) edge node[right] {j+=i;} (22);

  \node[draw,dotted,fit=(0) (7) (create),label=above:{main}] {};
  \node[draw,dotted,fit=(10) (12) (sum1),label=above:{t1}] {};
  \node[draw,dotted,fit=(20) (22) (sum2),label=above:{t2}] {};
\end{tikzpicture}

    }
    \caption{CFA for the functions of the program}
    \label{example-cfa}
  \end{minipage}
\end{figure}

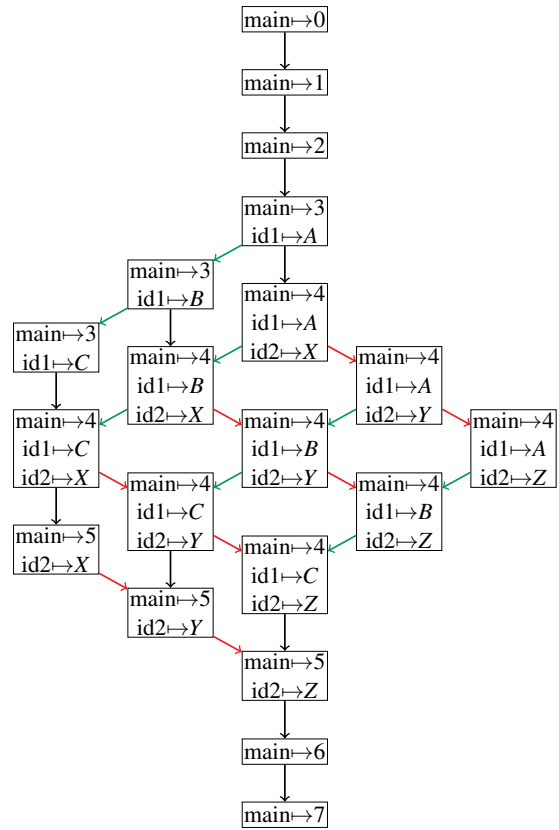
\begin{wrapfigure}{r}{0.5\textwidth}
  \begin{minipage}[b]{.48\textwidth}
    \centering
    \vspace{-12mm}
    \scalebox{.8}{


\tikzset{
  N/.style = { shape=rectangle, draw=black, inner sep=1pt,
               minimum height=12pt, minimum width=25pt}
}
\begin{tikzpicture}[x=1.9cm,y=1.05cm]
  \def\pthread{} 
  \def\shift{-1}
  \def\diamondblk{-.8}
  \def\endblk{-1.4}
  \foreach \x in {0,1,2} {
    \node[N] at (0,-\x) (\x) {$\text{main} {\mapsto} \x$};
  }
  \node[N] at (0,-3.2) (3) [align=center]{$\text{main} {\mapsto} 3$ \\ $\text{id1} {\mapsto} A$};

  \node[N] (4) at (-1,-4.2) [align=center]{$\text{main} {\mapsto} 3$ \\ $\text{id1} {\mapsto} B$};
  \node[N] (5) at (1+\shift,-4+\diamondblk) [align=center]{$\text{main} {\mapsto} 4$ \\ $\text{id1} {\mapsto} A$ \\ $\text{id2} {\mapsto} X$} ;

  \node[N] (6) at (-2,-5.2) [align=center]{$\text{main} {\mapsto} 3$ \\ $\text{id1} {\mapsto} C$};
  \node[N] (7) at (0+\shift,-5+\diamondblk) [align=center]{$\text{main} {\mapsto} 4$ \\ $\text{id1} {\mapsto} B$ \\ $\text{id2} {\mapsto} X$};
  \node[N] (8) at (2+\shift,-5+\diamondblk) [align=center]{$\text{main} {\mapsto} 4$ \\ $\text{id1} {\mapsto} A$ \\ $\text{id2} {\mapsto} Y$};

  \node[N] (9) at (-1+\shift,-6+\diamondblk) [align=center]{$\text{main} {\mapsto} 4$ \\ $\text{id1} {\mapsto} C$ \\ $\text{id2} {\mapsto} X$};
  \node[N] (10) at (1+\shift,-6+\diamondblk) [align=center]{$\text{main} {\mapsto} 4$ \\ $\text{id1} {\mapsto} B$ \\ $\text{id2} {\mapsto} Y$};
  \node[N] (11) at (3+\shift,-6+\diamondblk) [align=center]{$\text{main} {\mapsto} 4$ \\ $\text{id1} {\mapsto} A$ \\ $\text{id2} {\mapsto} Z$};

  \node[N] (12) at (-2,-7+\endblk) [align=center]{$\text{main} {\mapsto} 5$ \\ $\text{id2} {\mapsto} X$};
  \node[N] (13) at (0+\shift,-7+\diamondblk) [align=center]{$\text{main} {\mapsto} 4$ \\ $\text{id1} {\mapsto} C$ \\ $\text{id2} {\mapsto} Y$};
  \node[N] (14) at (2+\shift,-7+\diamondblk) [align=center]{$\text{main} {\mapsto} 4$ \\ $\text{id1} {\mapsto} B$ \\ $\text{id2} {\mapsto} Z$};

  \node[N] (15) at (-1,-8+\endblk) [align=center]{$\text{main} {\mapsto} 5$ \\ $\text{id2} {\mapsto} Y$};
  \node[N] (16) at (1+\shift,-8+\diamondblk) [align=center]{$\text{main} {\mapsto} 4$ \\ $\text{id1} {\mapsto} C$ \\ $\text{id2} {\mapsto} Z$};

  \node[N] (17) at (0,-9+\endblk) [align=center]{$\text{main} {\mapsto} 5$ \\ $\text{id2} {\mapsto} Z$};
  \node[N] (18) at (0,-10+\endblk-.2) [align=center]{$\text{main} {\mapsto} 6$};
  \node[N] (19) at (0,-11+\endblk-.2) [align=center]{$\text{main} {\mapsto} 7$};

  \path[->, thick] (0) edge node[right] {} (1);
  \path[->, thick] (1) edge node[right] {} (2);
  \path[->, thick] (2) edge node[right] {} (3);
  \path[->, thick] (3) edge node[right] {} (5);
  \path[->, thick] (4) edge node[left] {} (7);
  \path[->, thick] (6) edge node[left] {} (9);
  \path[->, thick] (9) edge node[left] {} (12);
  \path[->, thick] (13) edge node[right] {} (15);
  \path[->, thick] (16) edge node[right] {} (17);
  \path[->, thick] (17) edge node[right] {} (18);
  \path[->, thick] (18) edge node[right] {} (19);

  \foreach \x/\y in {3/4,4/6,5/7,7/9,8/10,10/13,11/14,14/16} {
    \path[->, thick, ForestGreen] (\x) edge node[right] {} (\y);
  }
  \foreach \x/\y in {5/8,8/11,7/10,10/14,9/13,13/16,12/15,15/17} {
    \path[->, thick, Red] (\x) edge node[right] {} (\y);
  }
\end{tikzpicture}

    }
    \caption{ARG of the interleaved threads of the program}
    \label{example-arg}
    \vspace{-8mm}
  \end{minipage}
\end{wrapfigure}

The following example applies our new \threadingcpa to a given program.
In contrast to the simplified illustration below,
a real-world analysis would combine the \threadingcpa with another analysis,
e.g., to track assignments, such as value analysis or BDD analysis.

The example program (cf. Fig.~\ref{example-code} for the source code) creates two additional threads
that change the value of global variables.
Afterwards, the main method checks the assignment of a global variable.
In this example, the property holds.
The program's functions are represented as CFAs in Figure~\ref{example-cfa}.
The \threadingcpa produces the ARG in Fig.~\ref{example-arg},
where each abstract state is labeled with the indices of the program locations of all active threads.

The analysis starts at entry location $l_0$ of the main function and analyzes all possible interleavings.
After reaching the statement \pthreadcreate,
an additional program location is tracked for the newly created thread,
e.g., when reaching program location~$l_3$ in the main function,
the abstract state is enriched with
the initial program location~$l_A$ of the newly created thread.

As the \threadingcpa merges its abstract states 
when reaching the same program locations via different execution paths,
the diamond-like structure in the ARG is the result
of interleaved thread-execution of two (or more) threads.
When exploring the statement \pthreadjoin,
the program-exit location of the exiting thread
is removed from the abstract state.
This is visible in Fig.~\ref{example-arg} for each abstract state
with an outgoing edge leading from program location~$l_4$
towards program location~$l_5$,
because the program-exit location~$l_C$ of the joining thread~$t_1$
(identified by id~$1$) is removed from the abstract state.


\section{Optimization}
\label{optimizations}

The simple definition of the \threadingcpa
allows (and needs) a wide range of optimization
to gain competitive efficiency.
In the following, we define some approaches and
show how fluently they match existing concepts in \cpachecker.

\subsection{Partitioning of Reached Abstract States}
\label{opt:partitioned-reachedset}

The reachability algorithm~\cite{CPA} has two important operators $merge$ and $stop$
that are defined as operations on sets of reached abstract states.
These operations can merge abstract states and combine their information into a new abstract state or detect coverage, i.e., an abstract state is implied by another one and thus the exploration can stop at that point.
In each iteration of the reachability algorithm,
these operators are by default applied to \emph{all} combinations of new explored abstract states and previously reached abstract states.
However, applying such an operator to \emph{all} previously reached abstract states is inefficient,
because most of the abstract states are irrelevant for a concrete application of these operators.
For example,
comparing abstract states from different program locations is useless,
because there will not be any important relation between them.

\emph{Partitioning} the set of abstract states
makes it possible to perform both operations much more efficiently,
as only a (small) subset of the previously reached abstract states has to be considered in the computation.
This basic optimization is also applied for verifying single-threaded programs.
Each partition is identified by a constant key
that is based on the program location of the abstract state,
as only states from equal program locations are considered for merging or coverage.
We extended the existing partitioning of abstract states,
such that it uses the tuple of program locations
for all threads in an abstract state.
This new partitioning can also be combined with partitionings
provided by other CPAs.

\subsection{Waitlist Order}
\label{opt:waitlist-order}

For finding property violations it is often sufficient
to only analyze interleavings with a low number of thread interleavings.
As the exploration algorithm in \cpachecker analyzes the reachable state space state by state,
there exists the possibility to \emph{prioritize abstract states} during the exploration:
The abstract states waiting to be analyzed are simply sorted by some criteria.
This optimization is a heuristic depending on
the internal structure of the analyzed program and the executed analysis.
For a bug-free program this heuristic does not bring any benefit.
however an existing error path in a faulty program might be found sooner.

The most-often used orderings of abstract states cause the state-space exploration
to perform either depth-first search (DFS) or breadth-first search (BFS),
i.e., the list of waiting abstract states is ordered
in the same manner as abstract states are explored (BFS) or reverse (DFS).
For multi-threaded programs, we added a new ordering of this list
based on the number of active threads,
such that states with fewer active threads are considered first.
The new ordering can also be combined with existing orderings,
i.e., the first criteria for ordering is based on the number of active threads,
the second criteria uses the exploration order.

\subsection{Partial-Order Reduction}
\label{opt:partial-order-reduction}

With multi-threaded programs, the most common form of optimization is \emph{partial-order reduction}~(POR)~\cite{Godefroid96POR, Valmari89POR, Peled93POR}.
POR aims to avoid unnecessary interleavings of threads
and improves the performance of the analysis by reducing the explored state space.
However, its application depends on the property to be verified,
because all necessary program paths must remain reachable.

In our case (reachability analysis), we started
with a simple separation of program operations (modeled as CFA edges)
into \emph{thread-local} and \emph{global} operations.
We conservatively apply a static analysis for all program variables and memory accesses,
on whether they are declared and used in \emph{global} scope or
only $locally$ in the context of a thread.
Because \cpachecker uses several dummy operations
(e.g., for temporary variables or function returns),
a majority of CFA edges is marked as \emph{thread-local}.

If a statement is \emph{thread-local} for a thread,
we do not simulate any interleaving after analyzing this operation,
but the analysis executes the current thread further,
until a global operation (in the same thread) is reached.
This behavior is \emph{sound}, because no interaction between threads is possible,
due to the definition of \emph{thread-local} operations.
Thus, we only need to synchronize all available threads after the next \emph{global} transition.

Our approach can analyze program with loops as well,
because we execute both paths,
i.e., the loop and the concurrent thread,
and none of them disables the other path.
Thus, any possible interaction
between CFA edges of the loop
and other threads is considered.
Our approach handles loops implicitly,
thus we do not have to actively check for loops,
but simply apply the reachability algorithm
combined with the described POR technique.


\section{Extensions}

During our work on the analysis of multi-threaded programs,
we explored some assumptions in \cpachecker
that need to be considered when integrating
such a basic analysis as the \threadingcpa.
We also noticed several features
that can also be specified or implemented for the analysis of multi-threaded programs.
In the following, we describe the extensions that we have developed
in order to use the full potential of the framework.

\subsection{Cloning for CFAs}

\cpachecker has a modular structure,
such that many components can be combined without knowing (and depending on) details about each other.
As the analysis of multi-threaded programs should fit into this design,
we decided not to modify each analysis
that should be combined with our new approach,
but use an approach that allows us
to re-use as much existing code as possible.

The basic problem with the existing components of \cpachecker is
that many of them rely on knowing only their current function scope,
and solely identify a variable by its name combined with
the name of the function scope it was declared in.
For example, many analyses (including value analysis and BDD analysis)
use the identifier $f{::}x$ for a variable $x$ declared in function $f$.
This identifier is used in the internal data structures
whenever the variable is used during the program analysis.
In a multi-threaded program,
the same function $f$ might be called in different threads,
such that $f{::}x$ is not unique for one variable any more
at a certain point in the program's execution.
The existing analyses do not know about two variables with the same identifier
and would, e.g., assign a wrong value to one of them.

Our solution is simple:
We use different function names for each thread
by \emph{cloning} the function and
inserting the corresponding indexed function name.
For a function $f$ we create a clone $f'$
by copying the corresponding CFA nodes from $L$ and edges from $G$,
while renaming all appearances of the function's identifier in the clone.
%
Cloning functions causes all function-local variables to be unique for different threads in the later applied analysis,
e.g., the identifier $f{::}x$ is distinct from $f'{::}x$.
An analysis using the identifier does not even have to know
whether the function is cloned
and can simply assume uniqueness of identifiers for all variables.

\subsection{Deadlock Detection}

A deadlock~\cite{IsloorM80Deadlock} is defined as an abstract state 
where two (or more) competing actions wait for each other to finish,
and thus neither ever does.
\cpachecker allows the user to define the goal of an analysis
by giving a specification in form of an automaton.
Detecting deadlocks in the program can be done by giving an observer automaton
that monitors the abstract states of the \threadingcpa
and reports deadlocks.
This approach is independent of any further analysis
and can be combined with, e.g., value analysis or BDD analysis.



\subsection{Mutex Locks}
\label{opt-mutex-locks}

Mutex locks are commonly used to synchronize threads,
e.g., to manage access to shared memory.
In our implementation, mutex locks are stored as part of the abstract state of the \threadingcpa.
If a mutex lock is requested along a CFA edge,
but not available in the preceding abstract state,
the transfer relation does not yield a successive abstract state for the CFA edge.

Additionally, we use mutex locks for more use cases:
We simulate atomic sequences of statements
and some aspects of partial order reduction
as mutex locks in the \threadingcpa.
Entering an atomic sequence requires an \emph{atomic mutex lock},
which is released after leaving the atomic sequence.
Consecutive CFA edges containing only thread-local operations
(see Section \ref{opt:partial-order-reduction})
are modeled and analyzed as atomic sequence.

%
%
%

\section{Evaluation}

In this section we evaluate different configurations of the \threadingcpa
and compare it with other state-of-the-art tools.
%
%
The evaluation is performed on machines with a 2.6\,GHz Intel Xeon E5-2650 v2 CPU
running Ubuntu 16.04 (Linux 4.4.0).
Each single verification run is limited to \SI{15}{min} of run time and \SI{15}{GB} of memory.
The \num{1016} benchmark tasks are taken from the category of multi-threaded programs
at SV-COMP'16~\footnote{\url{https://github.com/sosy-lab/sv-benchmarks/releases/tag/svcomp16}}.
The tasks are C programs,
where reaching a specific function call is considered as property violation.
%
We use \cpachecker~\footnote{\url{https://cpachecker.sosy-lab.org/}} 1.6.1 in revision \num{23011}.

\vspace{10mm}

\begin{figure}
  \begin{minipage}[b]{.48\textwidth}


\begin{tikzpicture}
\begin{semilogyaxis}[
  width=.94\textwidth, height=\textwidth,
  /pgfplots/table/y index=3, /pgfplots/table/header=false,
  xlabel=n-th fastest result, ylabel=CPU time (\second),
  xmin=0, xmax=1020, ymin=1, ymax=1000, mark repeat=150,
  legend entries={plain value analysis, + partitioning, + waitlist order, + POR (opt. VA)},
  every axis legend/.append style={at={(1,0)}, anchor=south east,
                                   outer xsep=0pt, outer ysep=0pt,
                                   nodes={right}, draw=none, fill=none},
  ]
  \addplot[solid, red, mark=diamond*, every mark/.append style={solid, fill=red}]
      table {benchmark/results.quantile.value-concurrency-noPartioning-DFS-noPOR.csv};
  \addplot[solid, blue, mark=square*, every mark/.append style={solid, fill=blue}]
      table {benchmark/results.quantile.value-concurrency-DFS-noPOR.csv};
  \addplot[solid, brown, mark=triangle*, every mark/.append style={solid, fill=brown}]
      table {benchmark/results.quantile.value-concurrency-noPOR.csv};
  \addplot[solid, ForestGreen, mark=otimes*, every mark/.append style={solid, fill=ForestGreen}]
      table {benchmark/results.quantile.value-concurrency.csv};

\end{semilogyaxis}
\end{tikzpicture}

    \caption{Quantile plot for different configurations
             of the value analysis,
             corresponding to step-wise applied optimization steps}
    \label{plot-value}
  \end{minipage}
  \hfill
  \begin{minipage}[b]{.48\textwidth}


\begin{tikzpicture}
\begin{semilogyaxis}[
  width=.94\textwidth, height=\textwidth,
  /pgfplots/table/y index=3, /pgfplots/table/header=false,
  xlabel=n-th fastest result, ylabel=CPU time (\second),
  xmin=0, xmax=1020, ymin=1, ymax=1000, mark repeat=150,
  legend entries={BDD analysis, interval analysis, opt. VA},
  every axis legend/.append style={at={(1,0)}, anchor=south east,
                                   outer xsep=0pt, outer ysep=0pt,
                                   nodes={right}, draw=none, fill=none},
  ]
  \addplot[solid, blue, mark=triangle*, every mark/.append style={solid, fill=blue}]
      table {benchmark/results.quantile.bdd-concurrency.csv};
  \addplot[solid, brown, mark=square*, every mark/.append style={solid, fill=brown}]
      table {benchmark/results.quantile.interval-concurrency.csv};
  \addplot[solid, ForestGreen, mark=otimes*, every mark/.append style={solid, fill=ForestGreen}]
      table {benchmark/results.quantile.value-concurrency.csv};

\end{semilogyaxis}
\end{tikzpicture}

    \caption{Quantile plot for different abstract domains
            using the \threadingcpa within \cpachecker}
    \label{plot-analyses}
  \end{minipage}
\end{figure}

\subsection{Optimization Steps}

First,
we show the effect of applying each optimization step from Section~\ref{optimizations} successively,
i.e., on top of the previous optimization.
Starting with a plain (non-optimized) configuration 
of the \threadingcpa combined with the value analysis,
we step-wise apply optimization in form of
\begin{itemize}
    \item reached-set partitioning (see Section~\ref{opt:partitioned-reachedset}) based on the abstract states,
    \item waitlist ordering (see Section~\ref{opt:waitlist-order}) based on the number of threads, and
    \item POR (see Section~\ref{opt:partial-order-reduction}) based on $local$-$scope$ and $global$ statements.
\end{itemize}

The optimization steps are independent of the value analysis
and can also be applied to any other analysis
like BDD analysis and interval analysis,
where the same benefit will be visible.
Figure~\ref{plot-value} shows a quantile plot
containing the run time of correctly solved verification tasks.
The evaluation shows that the verification process benefits from each of the optimization steps.
For small tasks that can be verified within a few second,
e.g., because of only a few thread interleavings in the program,
the benefit of optimization is small.
For tasks that need more run time the benefit becomes visible.

We noticed that the heuristic of ordering the waiting abstract states
is beneficial in two ways:
first, some property violations are found earlier
(some property violations need only a small number of interleavings);
second, some unsupported operations 
(like assigning several thread instances to the same thread identifier)
are reached earlier and the analysis can abort immediately
without wasting time.

Compared to the plain value analysis,
partitioning the reached set improves the performance and
reduces the run time of the analysis by more than an order of magnitude.
Additionally changing the waitlist order improves run time in several cases,
mostly for tasks with a property violation.
However, in our benchmark this optimization step does not lead to more correctly solved tasks.
POR causes a lower number of explored abstract states,
and thus the performance increases.

\subsection{Abstract Domains}

Second,
we combine the \threadingcpa with different analyses,
such as value analysis, interval analysis, and BDD analysis,
which are already implemented in the \cpachecker framework
and are normally used for the analysis of single-threaded programs.
We only evaluate the optimized version of each combination.
The analyses could also be combined with CEGAR~\cite{ClarkeCEGAR},
however the current benchmark does not benefit from it,
and thus we just execute a reachability algorithm to verify the specification.
We show that we can verify the majority of benchmark programs
and discuss strengths and weaknesses of the analyses.
As all compared analyses use the same framework (parser, algorithm, ...),
we expect our evaluation to be fair for all implemented approaches
and allow a precise comparison.
Figure~\ref{plot-analyses} shows the quantile plot of correct results
for the combinations of the \threadingcpa with other analyses.

%
The BDD analysis is optimized for BFS in the reachability algorithm,
whereas value analysis and interval analysis use DFS as basic order
for the list of waiting abstract states during the exploration algorithm (see Section~\ref{opt:waitlist-order}).
Thus, the state-space exploration traverses
program locations and thread interleavings in another order
and finds the corresponding abstract states in a different order, too.
Depending on the verification task,
this can result in an in- or decreased performance compared to the value analysis.

\begin{wrapfigure}{r}{0.5\textwidth}
  \begin{minipage}[b]{.48\textwidth}
    \vspace{20mm}


\begin{tikzpicture}
\begin{semilogyaxis}[
  width=.94\textwidth, height=\textwidth,
  /pgfplots/table/y index=3, /pgfplots/table/header=false,
  xlabel=n-th fastest result, ylabel=CPU time (\second),
  xmin=0, xmax=1020, ymin=1, ymax=1000, mark repeat=150,
  legend entries={CBMC, VVT, opt. VA},
  every axis legend/.append style={at={(1,0)}, anchor=south east,
                                   outer xsep=0pt, outer ysep=0pt,
                                   nodes={right}, draw=none, fill=none},
  ]
  \addplot[solid, red, mark=triangle*, every mark/.append style={solid, fill=red}]
      table {benchmark/results.quantile.cbmc.csv};
  \addplot[solid, blue, mark=diamond*, every mark/.append style={solid, fill=blue}]
      table {benchmark/results.quantile.vvt.csv};
  \addplot[solid, ForestGreen, mark=otimes*, every mark/.append style={solid, fill=ForestGreen}]
      table {benchmark/results.quantile.value-concurrency.csv};
\end{semilogyaxis}
\end{tikzpicture}

    \caption{Quantile plot for comparison of other verifiers
             with support for multi-threaded programs}
    \label{plot-other}
    \vspace{-8mm}
  \end{minipage}
\end{wrapfigure}
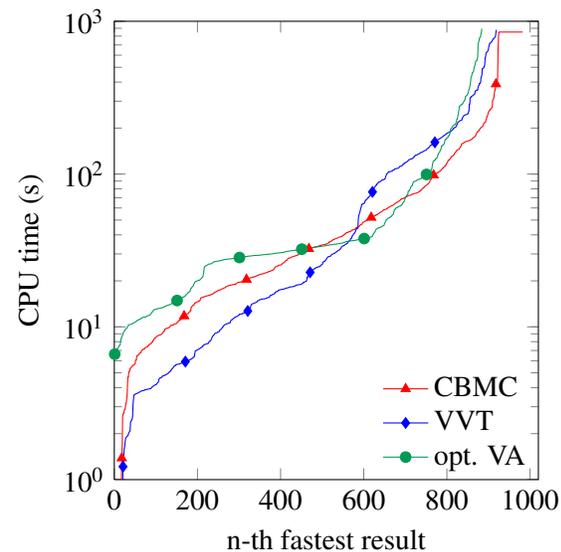
~

\subsection{Other Tools}

Third,
we compare the (optimized) value analysis with two other state-of-the-art verification tools,
namely \cbmc~\footnote{\url{http://www.cprover.org/cbmc/}} and \vvt~\footnote{\url{https://vvt.forsyte.at/}}.
Both tools are executed as in the SV-COMP'16
and are chosen, because they do not apply special approaches like sequentialisation,
but rely on a similar state-space exploration technique as our approach in \cpachecker.
%
Figure~\ref{plot-other} shows the quantile plot of correct results
for \cbmc, \vvt, and \cpachecker (using the optimized value analysis).
The \threadingcpa (combined with value analysis) is competitive with the other tools.
The plot for \cpachecker matches the trend of the other tools
with only some differences.
%
%
At the left side of the plot the initial start-up time of a few seconds for \cpachecker is visible,
whereas other tools already solve some of the given instance within this time.
%
Due to the missing support for pointer aliasing and array computations in the value analysis
as well as due to our simple kind of POR,
\cpachecker can not solve as many verification tasks as other tools
within the time limit.
%

\section{Conclusion}


This paper presents a basic approach to support
the analysis of multi-threaded programs in \cpachecker.
We formally defined a new \threadingcpa in the framework and 
demonstrated that several core components can be reused.
Re-using existing analyses is possible without any further overhead.
%
%
Due to our simple approach,
there are a few limitations that have to be considered
when verifying multi-threaded programs with \cpachecker.
%
%
Our approach for partial order reduction is simple
and can be extended with more advanced techniques
to further reduce the number of explored abstract states.
%
%
The maximum number of threads is bounded,
because of possible conflicts in function names.
To avoid naming conflicts,
we clone each function's CFA several times before starting the analysis.
The number of clones cannot be changed afterwards.
If we run out of clones during the analysis and would need more due to a naming conflict,
we abort the analysis and report an insufficient number of threads.


As the \threadingcpa identifies each thread only by the variable it is assigned to,
we currently can not analyze more complex thread management
such as pointer aliasing for the thread identifier or more complex locking mechanisms.
Our framework already contains a mechanism for exchanging information 
between abstract states on a state-level during the analysis.
The analysis of multi-threaded programs could be extended
to exchange information about thread management
with another analysis capable of such data,
such that we could analyze more difficult thread management with the \threadingcpa.

Possible ideas for optimization have been implemented and evaluated.
The evaluation shows that the results of different analyses based on the \threadingcpa
are competitive with other state-of-the-art tools.

\bibliography{sw,dbeyer}

\end{document}